\documentclass[12pt]{article}  
\usepackage{times}
\usepackage[numbers,sort]{natbib}
\usepackage{epsfig, amssymb, amsmath,multicol}

\usepackage{graphicx} 
\usepackage{wrapfig} 

\usepackage{geometry}
\usepackage[layout=modern]{advancedcoverpage}

\newcommand{\fb}[1]{
  \begin{center}
    \fbox{\parbox{.85\textwidth}{#1}}
  \end{center}
}

\def\msig{$M_{\rm BH}-\sigma_*$}

\newcommand{\msun}{M_{\odot}} 
\newcommand{\etal}{\textit{et al.}~}

\newcommand {\apgt} {\ {\raise-.5ex\hbox{$\buildrel>\over\sim$}}\ }
\newcommand {\aplt} {\ {\raise-.5ex\hbox{$\buildrel<\over\sim$}}\ }

\newcommand{\mbh}{M_{\rm BH}}

\def\simless{\mathbin{\lower 3pt\hbox {$\,\rlap{\raise
        5pt\hbox{$\char'074$}}\mathchar"7218\,$}}} 
\def\simgreat{\mathbin{\lower 3pt\hbox {$\,\rlap{\raise
        5pt\hbox{$\char'076$}}\mathchar"7218\,$}}} 

		\title{\Large Probing Quiescent Massive Black Holes: \\Insights from Tidal Disruption Events}
		\subtitle{\large A Whitepaper Submitted to the Decadal Survey Committee}
		\presentedto{\underline{Science Frontier Panels:}\\{\it Galaxies Across Cosmic Time (GCT)}}
		\author{\emph{\underline{Authors}}\\Suvi Gezari (Johns Hopkins, Hubble Fellow), Linda Strubbe, Joshua S. Bloom (UC Berkeley),  J. E. Grindlay, Alicia Soderberg, Martin Elvis (Harvard/CfA), Paolo Coppi (Yale), Andrew Lawrence (Edinburgh), Zeljko Ivezic (University of Washington), David Merritt (RIT), Stefanie Komossa (MPG), Jules Halpern (Columbia), and Michael Eracleous (Pennsylvania State)}
		\organization{\underline{Projects/Programs Emphasized:}\\1. The Energetic X-ray Imaging Survey Telescope (EXIST); {\tt http://exist.gsfc.nasa.gov}
                \\2. The Wide-Field X-ray Telescope (WFXT); {\tt http://wfxt.pha.jhu.edu}\\
  	          3.  Panoramic Survey Telescope \& Rapid Response System (Pan-STARRS); {\tt http://pan-starrs.ifa.hawaii.edu/public/}\\
                  4.  The Large Synoptic Survey Telescope (LSST); {\tt http://lsst.org}
                \\5.  The Synoptic All-Sky Infrared Survey (SASIR); {\tt http://sasir.org}}
		\date{}

\begin{document} 
\maketitle
\newpage

\fb{\large \bf \underline{Key Questions:}\\ \phantom{}\smallskip 1. What is the assembly history of massive black holes in the universe?\\ \smallskip 2. Is there a population of intermediate mass black holes that are the primordial seeds of supermassive black holes?\\ \smallskip 3. How can we increase our understanding of the physics of accretion onto black holes? \\ \smallskip 4.  Can we localize sources of gravitational waves from the detection of tidal disruption events around massive black holes and recoiling binary black hole mergers?}

\bigskip

\section{Introduction}

Dynamical studies of nearby galaxies suggest that most if not all galaxies with a bulge component host a central supermassive black hole (SMBH), and that the bulge and BH masses are tightly correlated \cite{mag98,fm00,gbb+00,tgb+02,bgh05,gh07}.  This is referred to as the \msig\ relation, where the velocity dispersion ($\sigma_*$) of bulge stars is a proxy of the bulge mass.  {\bf Why the BH mass, which is concentrated on such small scales, should correlate with the bulge mass on such large scales is a deep puzzle}. In the standard Lambda cold dark matter ($\Lambda$CDM) scenario, bulges supposedly grow by mergers whereas BHs grow from accretion of stars and gas (and perhaps dark matter \cite{ost00}). Mergers induce gravitational perturbations that allow gas to rapidly infall \cite{hern89}, fueling BH growth and ensuring at some level concurrent growth on large and small scales \cite{km00}. In what appears to be an efficient feedback mechanism \cite{dish05}, if the MBH lights up as an active galactic nucleus (AGN), then the growth of the BH can be halted as energy is pumped back into the gas reservoir quenches further growth \cite{sr98,matt05,king03,catt04,schaw06,kav07}. Even in the absence of merger-induced growth, random stellar encounters should serve to continually feed SMBH growth \cite{zhr02,mp04}. As a third channel for growth, BHs may grow via mergers with other BHs.  As summarized by Treu et al.~\cite{twmb07}, in the absence of feedback, the slope of the \msig\ relation could evolve in either sense depending on the respective growth histories.  

Clearly, gaining an unbiased (or at least a differently biased) understanding of the demographics \cite{mf01} of BHs in galactic centers as a function of cosmic time is {\bf crucial for understanding the growth of structure in the universe}. Yet the direct detection of SMBHs in the centers of quiescent galaxies via stellar and gas dynamics is severely limited by our ability to spatially resolve the sphere of influence of the central BH, $R_{\rm sph} \sim G\mbh/\sigma_*^{2}$.   Thereby direct BH mass measurements are largely restricted to a limited number of nearby non-active systems for which accurate BH masses can be obtained from dynamical measurements \cite{sog+03,gsh+05,mmh+95,gmh+96,lo05} and active galaxies with intensive reverberation mapping campaigns \cite{pet93,gkh+00,ofm+04}.

\section{Tidal Disruption Events: a tool for probing black holes}

Stellar dynamical models predict that once every $10^{3}-10^{5}$ yr a star in a galaxy will pass within the tidal disruption radius of the central BH, $R_{p}<R_{T}=R_{\star}(\mbh/M_{\star})^{1/3}$, and will be torn apart by tidal forces \cite{mt99,wm04}.  After the star is disrupted, at least half of the debris is ejected from the system, and some fraction remains bound to the BH and is accreted.  The fallback of debris onto the BH produces a luminous electromagnetic flare that is expected to peak in the UV/X-rays and radiate close to the Eddington luminosity\cite{Rees1988,Ayal2000,Ulmer1999}.  

The detection of a tidal disruption event (TDE) is unambiguous evidence for
the presence of a central BH, and enables the detection of
quiescent SMBHs -- those not accreting, or at such low rates as to be
undetectable. TDEs also provide a cosmic laboratory for studying the detailed 
physics of accretion onto BHs.
The properties of the 
flare from the accreting stellar debris 
(luminosity, light curve, and spectral energy distribution) are dependent on the mass and 
spin of the BH.  {\bf Thus, detailed observations of tidal disruption 
flares can provide an independent means of measuring the masses and spins 
of dormant BHs in distant galaxies.}  The disruption of compact
objects, specifically white dwarfs, is also an interesting possiblity since
it can only occurr around BHs with masses of order $10^{5} \msun$ or less and should
have a distinct spectroscopic signature \cite{ses08}.  However, in the case
of the most massive BHs, the tidal disruption radius is 
smaller than the event horizon 
and stars are swallowed whole without disruption (for a solar-type star 
$M_{crit} \apgt 10^{8} \msun$, \cite{hills75}). 

\vspace{5pt}
\subsection{Progress to Date}

Although the volumetric rate of TDEs is low ($10^{-5}-10^{-7}$ yr$^{-1}$ Mpc$^{-3}$), 
about a dozen candidate TDEs have emerged in all-sky 
X-ray surveys (ROSAT and XMM-Newton Slew Survey) and the
\textsl{GALEX} Deep Imaging Survey in the UV.
  
The ROSAT All-Sky Survey detected large amplitude ($> 100$), luminous
($10^{42}-10^{44}$ ergs s$^{-1}$) soft X-ray outbursts from 5 galaxies
\cite{kom02} that showed no previous 
evidence of AGN
activity, and detection rates consistent with theoretical predictions
for the rates of TDEs \cite{donley}.  More recently the XMM-Newton Slew Survey \cite{esq06,esq08} 
detected two galaxies with large amplitude ($80-90$),
luminous ($10^{41}-10^{43}$ ergs s$^{-1}$) soft X-ray outbursts when
compared to their ROSAT upper limits.  Although two of these candidates (NGC 5905 and NGC 3599)
were found to have optical spectra with narrow emission lines indicative of a 
low-luminosity AGN \cite{gez03,esq08}, 
the large amplitudes of their X-ray flares were more extreme than any previously known cases of AGN variability.
These luminous, soft X-ray outbursts were
best explained as the result of a sudden increase in 
fuel supply to an otherwise quiescent
central BH due to the tidal disruption and accretion of a star. 

The X-ray spectra of the candidate TDEs detected by
ROSAT and XMM-Newton can be fitted with blackbody spectral energy
distributions (SEDs) with temperatures of $(6-12)\times10^{5}$ K, or 
extremely soft power-law spectra with $\Gamma = 3-5$, 
where $f_{E} \propto E^{-\Gamma}$.  
The Rayleigh-Jeans tail of this blackbody emission should be prominent in the UV, as
well as benefit from a strong negative K correction.  In a systematic search
of $\sim 3$ deg$^{2}$ of the \textsl{GALEX} Deep Imaging Survey in the FUV
and NUV two luminous ($> 10^{43}$ ergs s$^{-1}$)
flares were detected from galaxies at $z=0.37$ and $z=0.33$ with no signs of AGN activity in
their nuclei in the form of optical emission lines or hard X-ray emission 
\cite{gez06,gez08}.

The simultaneous optical difference imaging light curves of the UV flares, extracted
from the CFHT Legacy Survey with 
an excellent cadence of days, are well described by the 
$t^{-5/3}$ power-law decay predicted both analytically and from 
numerical simulations to be the power-law decline of the mass 
fallback rate of debris from a tidally disrupted star 
\cite{Phinney90,EvansKochanek1989,Ayal2000,Lodato2008}.  
The UV/optical SEDs of the flares are fitted with blackbody temperatures of 
$\sim 5\times10^{4}$ K, or power-laws with $\alpha=+1.0-+1.5$, where $f_{\nu} \propto \nu^{\alpha}$.   
Soft X-ray emission was also detected during the flares, consistent with 
a second blackbody component with $T_{BB}=(2-5) \times  10^{5}$K.  
In Gezari \etal (2008) \cite{gez08}, they run a Monte Carlo simulation of their survey efficiency and find
that the observed detection rate of the UV/optical candidates is consistent with the black-hole-mass dependent TDE rates from stellar dynamical models from Wang \& Merritt (2004) \cite{wm04} if the assumption is made that the fraction of flares that radiate at the Eddington luminosity decreases with increasing BH mass \cite{Ulmer1999}, and that stars are spun-up to near break up when disrupted \cite{li02}.  

\vspace{5pt}
\section{Future Prospects}

The detection of the light curves of tidal disruption events for the first time in the optical has encouraging implications for the next wave of high-cadence, optical/infrared time-domain surveys, which will be covering a large enough volume to detect tens of tidal disruption events per year.  
The detailed optical light curves and colors measured from high cadence difference imaging will be able identify tidal disruption events in the nuclei of galaxies, and trigger multiwavelength follow-up imaging and spectroscopy.

Strubbe \& Quataert (2009) \cite{sq09} calculate models of tidal disruption flares at optical wavelengths in order to predict photometric and spectroscopic diagnostics.  In these models, about half of the stellar debris accretes in a thick disk that radiates as a multicolor blackbody peaking around 0.1 keV.  For $\sim 1$ month to $\sim 1$ year, the accretion rate remains high enough for the disk to shine at close to the Eddington luminosity.  The other half of the stellar debris escapes from the BH, spreading into an angle that is larger for lower BH masses.  This material is irradiated by the disk, and re-emits at longer wavelengths via recombination, bremsstrahlung, and lines.  For BHs having $M_{\rm BH} \sim 10^7 M_\odot$, the optical emission is dominated by the Rayleigh-Jeans tail of the disk's emission, and is very blue.  For $M_{\rm BH} \sim 10^5-10^6 M_\odot$, reprocessing by the unbound material can boost and redden the optical emission significantly.  Super-Eddington rates of infall may also drive a wind that could boost and redden the optical emission even more.

In Figure 1, we use these emission models to predict detection rates with upcoming surveys.  We assume a rate of $10^{-5} {\,\rm yr}^{-1}$ per BH based on ROSAT detections \cite{donley} and conservative theoretical estimates, and assume it is independent of BH mass for simplicity.  We use the BH mass function of Tundo et al. (2007) \cite{tundo07}, and assume it remains at $10^{-2} {\rm Mpc}^{-3}$ down to $10^5 M_\odot$.  Our calculations include cosmological effects.  We plot rates in the $g$ and $i$ band for the Pan-STARRS 1 Medium Deep Survey (PS1 MDS) and $3\pi$ Survey, and for LSST.  Detection rates peak at $M_{\rm BH} \sim {\rm few} \times 10^7 M_\odot$, but include a significant number of intermediate mass black holes (IMBHs) as well, particularly at $i$-band.  With these assumptions, the MDS will detect $0.4 {\,\rm yr}^{-1}$, $3\pi$ will detect $13 {\,\rm yr}^{-1}$, and LSST will detect $130 {\,\rm yr}^{-1}$. This is in agreement with the predictions for Pan-STARRS and LSST in Gezari \etal (2008) \cite{gez08} when we use the black-hole-mass dependent TDE rate used in their calculation \cite{wm04}.
Dust obscuration of the central regions of some galaxies might hide an appreciable fraction of events.  
Therefore synoptic
searches in the infrared might be a useful complement to optical and X-ray studies. SASIR is a pre-phase A concept for a 
wide-field synoptic infrared survey with 
science operations expected at the end of the decade. Typical visits should yield depths of 
few $\mu$Jy in simultaneous YJHK imaging over 450 square degrees per night, 
with several thousands of square degrees that could be imaged on timescales relevant for TDE discovery.

While surveys like Pan-STARRS 1 will confirm the feasibility of using optical surveys alone to discover and identify TDEs, wider, deeper surveys like LSST will be the factory that will produce large enough samples of these 
events to study their ensemble properties, and map out their rates as a function of galaxy type and redshift.  
TDEs can even be a powerful tool for
studying the structure of galactic nuclei.  The rate of TDEs is very sensitive to a galaxy's nuclear stellar 
density profile \cite{mago99,wm04} and non-axisymmetric geometry \cite{mp04}, 
as well as the presence of a nuclear star cluster \cite{merr08}.

\medskip

\begin{minipage}{3in} {\small {\bf Fig 1.} {\small \it Predicted number of tidal disruption flare
   detections per year as a function of central BH mass for
   the Pan-STARRS Medium Deep (MDS), all-sky survey ($3 \pi$), and LSST survey. Adapted from Strubbe and Quataert (2009).}}
\end{minipage} \hskip 0.2in
\begin{minipage}{3.3in} \centerline{\epsfig{file=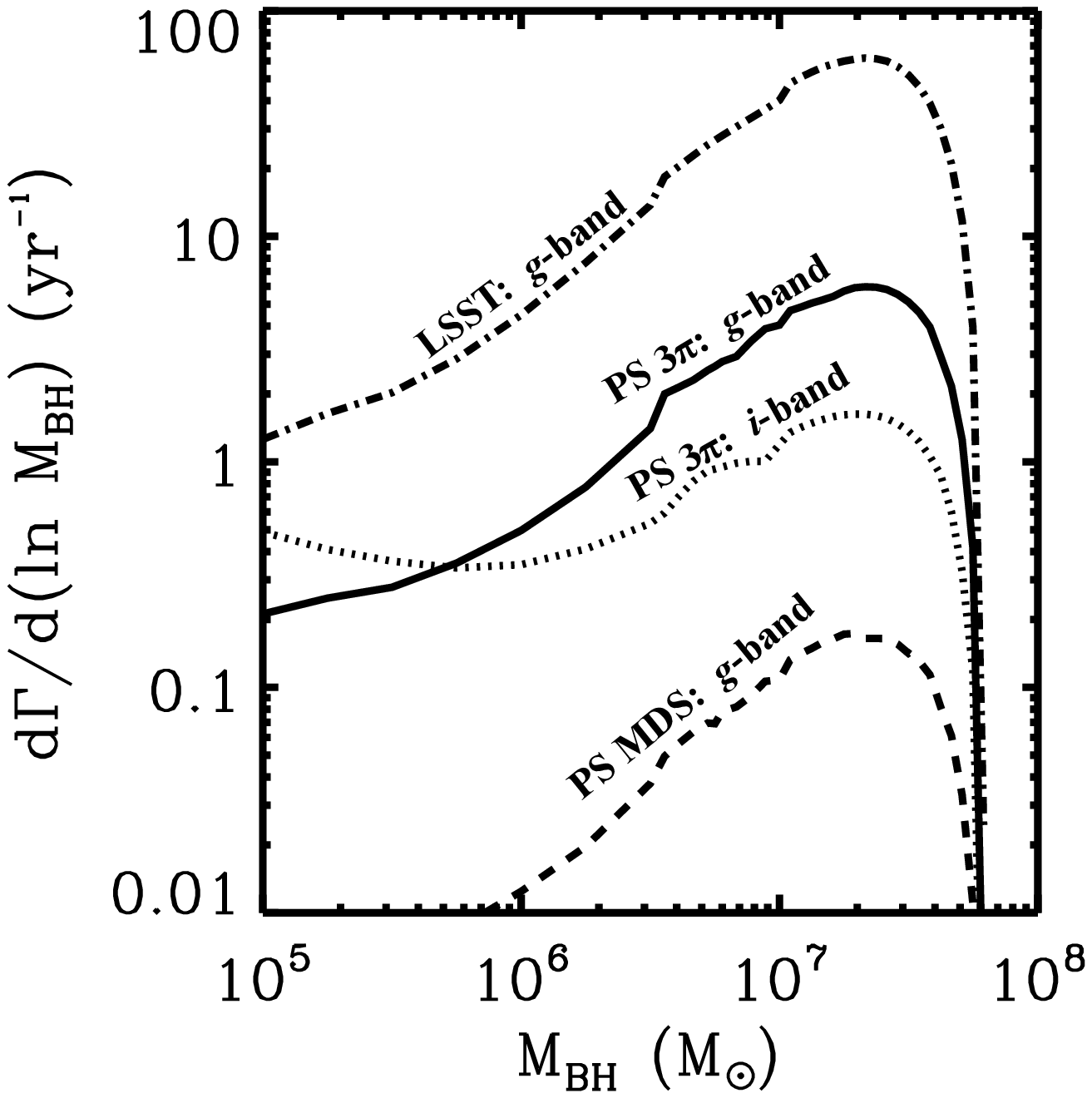,width=3.2in}}
\end{minipage} \hskip 0.1in

\medskip

\vspace{5pt}
\subsection{The Next Decade}

A more sophisticated theoretical treatment of TDEs than the invocation of simple fallback is clearly warranted.  Just what are the signatures at radio and X-ray wavebands is clearly not well understood ab initio --- and indeed the constraints placed across the electromagnetic spectrum will directly feed into and {\bf inform} models of accretion in MBHs.

It is clear that spectroscopic and multi-wavelength follow-up of optical detections will be  important.  Optical/UV spectroscopy and color evolution will help confirm the nature of an event, and constrain the BH mass and pericenter distance of the disrupted star.  These quantities are further indicated by the duration and bolometric luminosity of the flare, whose spectrum should peak at ultraviolet to soft X-ray wavelengths.  By analogy to X-ray binaries, tidal flares may become bright at radio wavelengths at late times when the fallback rate subsides below $\sim 10^{-2}\dot{M_{\rm Edd}}$, perhaps driving a jet \cite{remmc06}.  The characteristic radius of the emission derived from the spectral energy distribution (SED) of a flare may be used to probe the innermost stable circular orbit (r$_{\rm ISCO}$), which is strongly dependent on BH spin.

\medskip
\begin{minipage}{3in} 
\centerline{\epsfig{file=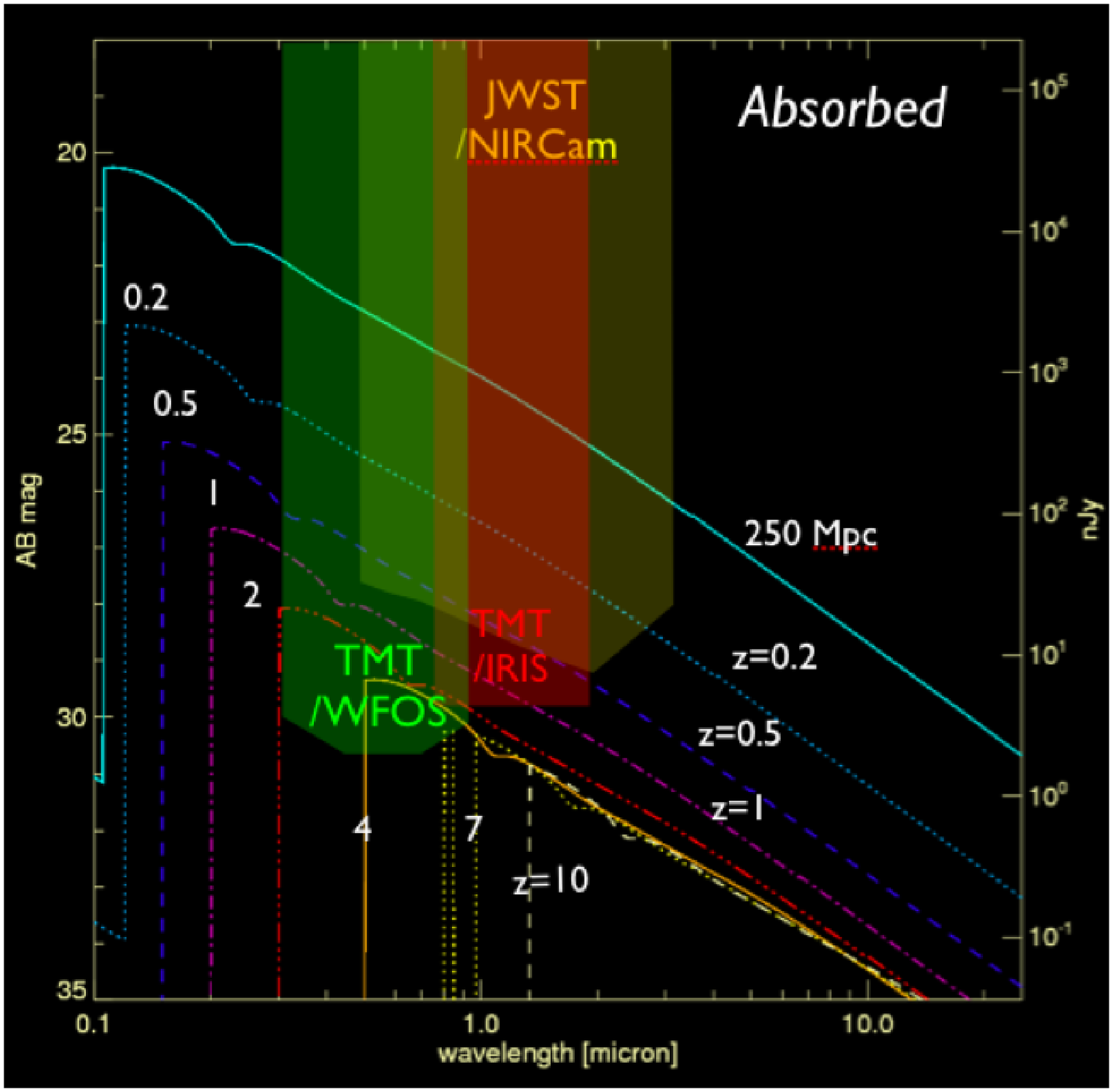,width=3in}}
\end{minipage} \hskip 0.1in
\begin{minipage}{3.4in} {\small {\bf Fig 2.} {\small \it Illustration of the observability of TDEs (near peak brightness) to high redshift using JWST and TMT. The strong negative correction ($K \approx -7.5 \log_{10} (1 + z)$) largely counteracts the $d_L$ dimming and dust obscuration.}}
\end{minipage}
\medskip

Figure 2 illustrates how measurements of the broadband SED and
detailed light curves of events could followed in the optical/NIR with instruments such
as JWST and TMT, which will be sensitive to the Rayleigh-Jeans tail of 
the emission out to high redshifts.  Indirect spectroscopic signatures of a 
TDE in the form of a ``light echo'' of fading high-ionization emission lines 
from the surrounding interstellar medium could be used as a diagnostic tool to study the energetics of
the flare \cite{kom08b}, while NIR imaging could be used to probe dusty gas in the cores of non-active
galaxies that has been heated to high temperatures.  

The proposed EXIST mission plans to conduct a very sensitive all-sky hard
X-ray time domain survey with a wide-field High Energy Telescope (HET) imaging 
the full sky every 3h in the $5-600$ keV band  
but with TDE detections likely in 
the $\sim5-20$ keV band. EXIST would have the unique ability to follow up
all events with an onboard optical-Infrared Telescope (IRT) and Soft X-ray
Imager (SXI).  The discovery rate by HET of TDEs in the hard X-rays
depends on the fraction of photons that are emitted at energies $> 5$ keV.
However, evidence for a hardening with time of the spectra of two
ROSAT TDE sources (RXJ1242.6 and NGC 5905) to
$\Gamma \sim 2.5$ \cite{kom99, kom04} points favorably towards the detection 
capabilities of EXIST \cite{gri04}.
Given that 30\% of the ROSAT and XMM-Newton TDE candidates had $\Gamma \le 3$ 
during the {\it peak} of the flare, and assuming that this is representative of the whole
population, the
BH mass function and tidal disruption rates described earlier
would imply $\sim 100$ detections per year by the HET.

A prompt (10 sec) flash of hard X-ray emission
has been proposed to occur as a result of the tidal compression of the star at the time of disruption \cite{kob04}.  Detection of this  X-ray flash by the EXIST HET would mark the time of disruption to high accuracy, and the subsequent emission from the fallback of the debris could be monitored with IRT and SXI.  Such precise measurements of the time of disruption and the shape of the power-law decay would place strict limits on the mass of the central BH and the mass of the star disrupted 
\cite{li02}.  

The proposed Wide-Field X-ray Telescope (WFXT) will survey the sky in the
soft X-ray band ($0.4-6$ keV).  While not explicitly a time domain survey,
the mission plan includes a deep survey over 100 deg$^{2}$ that will 
require multiple 1.5 ks visits to accumulate their deep exposure goal of 
395 ks.  This would enable the discovery of tens of TDEs in an energy
band that is expected to be close to the peak of their SEDs.  

{\bf We expect that by 2020, the systemic study of TDEs will be a major industry}. A large census of events will allow us to map the masses and spins of SMBHs in galaxies as a function of redshift and galaxy type, as well as use the measured rates to probe the stellar structure of the host galaxy nuclei.  This will allow the construction of an \msig\ relation entirely independent from methods currently available, allowing for unique constraints on the  central questions outlined concerning BH and galaxy growth.  Although the
detection sensitivity of the optical synoptic surveys shown in Figure 1 
are ``tuned'' to BH masses of a limited mass range, $(1-3) \times 10^{7} \msun$,
they will be a sensitive measure for deviations from the locally
established scaling relations between the mass of the BH and
their host galaxy properties.  
The low-mass tail of the detection sensitivity of LSST may even 
reveal the presence of IMBHs in dwarf galaxies 
and globular clusters \cite{ram08}.  There is also the exciting potential 
to use the electromagnetic detection of a TDE to localize 
the burst of gravitational waves detectable by LISA when a star with a compact core 
approaches the tidal disruption radius or during the partial
disruption of a white dwarf \cite{kob04,ses08,frei03}, and to detect and localize recoiling
SMBHs by TDEs from the cluster of stars which remains bound to them \cite{kom08}. 

\begin{multicols}{2}
\begin{scriptsize}
\bibliographystyle{astro2010}
\bibliography{journals_apj,astro2010-tdfs}
\end{scriptsize}
\end{multicols}
\end{document}